\documentclass[12pt]{article}
\usepackage[a4paper,left=2cm,right=1.2cm,top=3cm,bottom=4cm]{geometry}
\usepackage{amsmath,amstext, amsthm, empheq, extpfeil,  
amsbsy,amssymb,marvosym,fancyhdr,graphicx,amscd,amsfonts,latexsym,delarray,stackrel,lineno,color,cite,appendix,xcolor,url,hyperref,float}
\usepackage{wasysym}
\usepackage{subfig}

\usepackage{srcltx}
\usepackage{mathrsfs}
\usepackage{float} 
\usepackage[all]{xy}
\usepackage{t1enc}
\usepackage{pifont}
\usepackage{authblk}

\usepackage{mathpple}

\usepackage[T1]{fontenc}

\definecolor{Green}{rgb}{0,1,0}
\definecolor{Blue}{RGB}{0,0,191}
\definecolor{mathmodecolor}{RGB}{0,102,0}
\definecolor{keywordcolor}{RGB}{0,51,151}
\definecolor{sourcebackgroundcolor}{RGB}{255,247,223}
\definecolor{unixagred}{RGB}{255,0,0}
\definecolor{lightgray}{RGB}{191,191,191}
\definecolor{green}{RGB}{1,191,191}

\newcommand*\patchAmsMathEnvironmentForLineno[1]{%
  \expandafter\let\csname old#1\expandafter\endcsname\csname #1\endcsname
  \expandafter\let\csname oldend#1\expandafter\endcsname\csname end#1\endcsname
  \renewenvironment{#1}%
     {\linenomath\csname old#1\endcsname}%
     {\csname oldend#1\endcsname\endlinenomath}}%
\newcommand*\patchBothAmsMathEnvironmentsForLineno[1]{%
  \patchAmsMathEnvironmentForLineno{#1}%
  \patchAmsMathEnvironmentForLineno{#1*}}%
\AtBeginDocument{%
\patchBothAmsMathEnvironmentsForLineno{equation}%
\patchBothAmsMathEnvironmentsForLineno{align}%
\patchBothAmsMathEnvironmentsForLineno{flalign}%
\patchBothAmsMathEnvironmentsForLineno{alignat}%
\patchBothAmsMathEnvironmentsForLineno{gather}%
\patchBothAmsMathEnvironmentsForLineno{multline}%
}

\def\elel{(  \downarrow 1)}
\def\dodo{(   \downarrow 3)}

\def\higgs{{\bf H}}

\def\Aut{{\rm Aut}}

\def\Ker{{\rm Ker}}

\def\Tr{{\rm Tr}}

\def\C{{\mathbb C}}

\def\R{{\mathbb R}}
\def\Z{{\mathbb Z}}

\def\Tr{{\rm Tr}}

\def\cA{{\mathcal A}}

\def\cH{{\mathcal H}}

\def\cM{{\mathcal M}}

\def\dar[#1]{\ar@<2pt>[#1]\ar@<-2pt>[#1]}

\newcommand{\ie}{{\it i.e.\/}\ }

\newcommand{\cf}{{\it cf.}}

\newcommand{\nil}[1]{}
\newcommand{\noopsort}[1]{}

\makeatletter
\DeclareMathOperator{\exterior}{\@ifnextchar^\@exterior{\@exterior^{}}}
\def\@exterior^#1{\mathop{\bigwedge\nolimits^{\!#1}}}
\makeatother

\title{Noncommutativity and Physics: A non-technical review}
\author
      {Ali H. Chamseddine$^{1,4}$, Alain Connes$^{2,3}$ and Walter D. van
Suijlekom$^{5}$
      \\
      $^{1}$Physics Department, American University of Beirut, Lebanon\\
            $^{2}$College de France, 3 rue Ulm, F75005, Paris, France\\
      $^{3}$I.H.E.S. F-91440 Bures-sur-Yvette, France\\
         $^{4}$Ludwig Maximilian University, Theresienstrasse 37, 80333, Munich, Germany\\
            $^{5}$Institute for Mathematics, Astrophysics and Particle Physics, Radboud
University Nijmegen, Heyendaalseweg 135, 6525 AJ Nijmegen, The Netherlands.
\\
\bigskip
\texttt{chams@aub.edu.lb, alain@connes.org, waltervs@math.ru.nl}
      }

\begin{document}

\maketitle
\begin{abstract}
  We give an overview of the applications of noncommutative geometry to physics. Our focus is entirely on the conceptual ideas, rather than on the underlying technicalities. Starting historically from the Heisenberg relations, we will explain how in general noncommutativity yields a canonical time evolution, while at the same time allowing for the coexistence of discrete and continuous variables. The spectral approach to geometry is then explained to encompass two natural ingredients: the line element and the algebra. The relation between these two is dictated by so-called higher Heisenberg relations, from which both spin geometry and non-abelian gauge theory emerges. 
  Our exposition indicates some of the applications in physics, including Pati--Salam unification beyond the Standard Model, the criticality of dimension 4, second quantization and entropy.

\end{abstract}
\tableofcontents
\section{Introduction}
Our contribution to this volume on "noncommutativity and physics" will describe the key role of the transition from commutative to noncommutative algebra  starting from Heisenberg's discovery of matrix mechanics. The conceptual reason for the power of this transition is that the encoding by noncommutative algebra retains more information than its commutative counterpart. We are in fact very well acquainted with this fact when we write words.   Writing respects the order of the letters and this allows one to encode information in a very effective manner. Passing to the commutative ignores the order of letters and  equates words whose letter content is the same as happens in anagrams. The nuance between the noncommutative and the commutative is the same as in the game of scrabble where the same set of letters might correspond to quite different words. In a suggestive manner one can view quantization as the act of lifting from the commutative (semiclassical) shadow to the noncommutative real world. In Section \ref{secttime} we explain in which sense noncommutative spaces are dynamical, and possess a canonical time evolution.  
This fact is at the root  of noncommutative geometry, leaving the static case to ordinary geometry. In Section \ref{sectcalculus} we explain how noncommutativity of real variables is the key for the coexistence of discrete and continuous variables. They do coexist in the quantized calculus which gives a perfect stage for infinitesimal variables and allows for the  coexistence of the discrete with the continuum precisely by the noncommutativity of the actors (the operators) on this stage.
The irruption of noncommutativity in physics described above had a strong impact on mathematics which led to a reconstruction of geometry inside the Hilbert space formalism of quantum mechanics. 
 Encoding geometry in this formalism has two key ingredients, the line element $ds$ whose incarnation as the Dirac propagator is described in Section \ref{sectlocalact}, and the algebra of coordinates where once again noncommutativity enters and as explained in Section \ref{sectalgebra}  allows one to encode even ordinary spaces at a much lesser price than with commutative algebra.
We have  
understood the minimal amount of noncommutativity needed to obtain spin 4-manifolds,  
and this  lead to our Pati-Salam extension of the standard Model. The special role of dimension 4 is described in Section \ref{sectsobolev}. The non-abelian gauge theory are witnesses of the slight amount of noncommutativity required in the encoding of 4-manifolds.
 The action principle that leads to the Einstein-Hilbert gravitational action coupled with the above extension of the standard Model is the spectral action principle. It acquires the meaning of an entropy when working at the second quantized level, and we close this short paper in Section \ref{secondq} by explaining the potential role of second quantization for the spectral paradigm of geometry.

\subsection*{Acknowledgements}
The work of A. H. C is supported in part by the National Science Foundation Grant No. Phys-1912998 and the Arnold Sommerfeld Center at LMU.



\section{Noncommutativity generates time}\label{secttime}
The irruption of noncommutativity in physics can be traced back to this  night of June 1925 when, around three in the morning, W. Heisenberg while working alone in the Island of Helgoland in the north sea, discovered matrix mechanics and the noncommutativity of the phase space of microscopic mechanical systems. Until that day all the physics manipulations of observable  quantities were done within commutative algebra and the discovery that fundamental observables such as position and momentum could fail to obey the elementary commutative law is a fundamental  turning point which is also at the origin of the mathematical theory of noncommutative geometry. Indeed at the mathematical level it showed the relevance in geometry of spaces whose coordinates do not commute.   After the formulation of Heisenberg's discovery as matrix mechanics by Born and Jordan in their 1925 paper, von Neumann reformulated quantum mechanics using Hilbert space operators and went much further with Murray in identifying ``subsystems" of a quantum system as ``factorizations" of the underlying Hilbert space $\cH$. They discovered unexpected factorizations which did not correspond to tensor product decompositions $\cH=\cH_1\otimes \cH_2$ and developed the theory of factors which they classified into three types. Type I corresponds to factors $M$ associated to ordinary tensor product decompositions $\cH=\cH_1\otimes \cH_2$ and such factors are formed of those operators of the form $T\otimes 1$. Factors of type II are those which posses a trace and factors of type III are those which are neither of type I or II. The Tomita-Takesaki theory \cite{TT} extended to  factors the correspondence which exists in the type I case or in the type II case using the trace, between the Boltzmann-Gibbs state $\phi$ and the Heisenberg evolution $\sigma_t$ of observables, both expressed in terms of the Hamiltonian $H$
$$
\phi(A)=\Tr(A \exp(-\beta H))/\Tr(\exp(-\beta H))\to \sigma_t^\phi(A)=\exp(it H)a\exp(-it H)
$$ 
The starting point of the classification of factors and the reduction of type III to type II was the discovery in \cite{Co_2bis} that the evolution $\sigma_t^\phi\in \Aut(\cM)$ is in fact independent of the choice of the (faithful normal) state $\phi$ on the factor $\cM$ provided one divides the group $\Aut(\cM)$ of automorphisms of the factor  $\cM$ by those automorphisms $\alpha$ which exist as a trivial consequence of noncommutativity, i.e. those of the form, for some unitary $U\in \cM$,
$$
\alpha(A)=UAU^*, \forall A \in \cM
$$
Such automorphisms are called "inner" and they form a normal subgroup ${\rm Int}(\cM)\subset \Aut(\cM)$ of the group of automorphisms of $\cM$. 
This showed that factors $\cM$ admit a canonical time evolution \cite{Co_2bis,thesis} \ie a canonical homomorphism 
$$
\R \stackrel{\delta}{\longrightarrow} {\rm Out}(\cM)=\Aut(\cM)/{\rm Int}(\cM)
$$
by showing that the class of the modular automorphism $\sigma_t^\varphi$ in ${\rm Out}(\cM)$ does not depend on the choice of the faithful normal state $\varphi$.  The above uniqueness of the class of the modular automorphism \cite{Co_2bis}  drastically changed the status of the two invariants which had previously been introduced  in \cite{CoCR72a}, \cite{CoCR72b} by making them computable. The kernel of $\delta$,  
 $T(M)=\Ker \delta$ forms a subgroup of $\R$, the {\em periods} of $M$, and many non-trivial non-closed subgroups appear in this way. The fundamental invariant of factors is the {\em modular spectrum}  $S(M)$ of \cite{CoCR72a}. Its intersection $S(M)\cap \R_+^*$  is a closed subgroup of $\R_+^*$, \cite{acvd}, and this gave the subdivision of type III into 
 type  III$_\lambda$ $\iff$  $S(M)\cap \R_+^*=\lambda^Z$.  Subsequently, the classification of factors of type  III$_\lambda$, $\lambda\in [0,1)$ was shown to be reduced to that of type II and automorphisms $$
M=N\rtimes_\theta \Z, \ \ N \ \textbf{type} \ {\rm II}_\infty, \ \theta \in \Aut(N)
$$
In the case III$_\lambda$, $\lambda\in (0,1)$, $N$ is a factor and the automorphism $\theta \in \Aut(N)$ is of module $\lambda$ \ie it scales the trace by the factor $\lambda$.
In the III$_0$ case, $N$ has a non-trivial center and, using the restriction of $\theta$ to the center, this gave a very rich invariant, a flow, which was used in 1972 \cite{Cocras} to show the existence of hyperfinite non ITPFI factor.  
 Only the case III$_1$ remained open in \cite{thesis} and was solved later by Takesaki \cite{Tak} using crossed product by $\R_+^*$.

\section{Discrete and continuous variables}\label{sectcalculus} 

One of the really new totally unexpected features of quantum
mechanics is the fact that outcomes of microscopic experiments can not be repeated. Even if you consider just a single
slit experiment and you shoot electrons or photons through a slim slit of size comparable to the wavelength of the particles,
the exact location where the particle will land on a target on the other side of the slit is something that cannot be reproduced.

\begin{figure}[H]
  \centering
  \includegraphics[scale=.6]{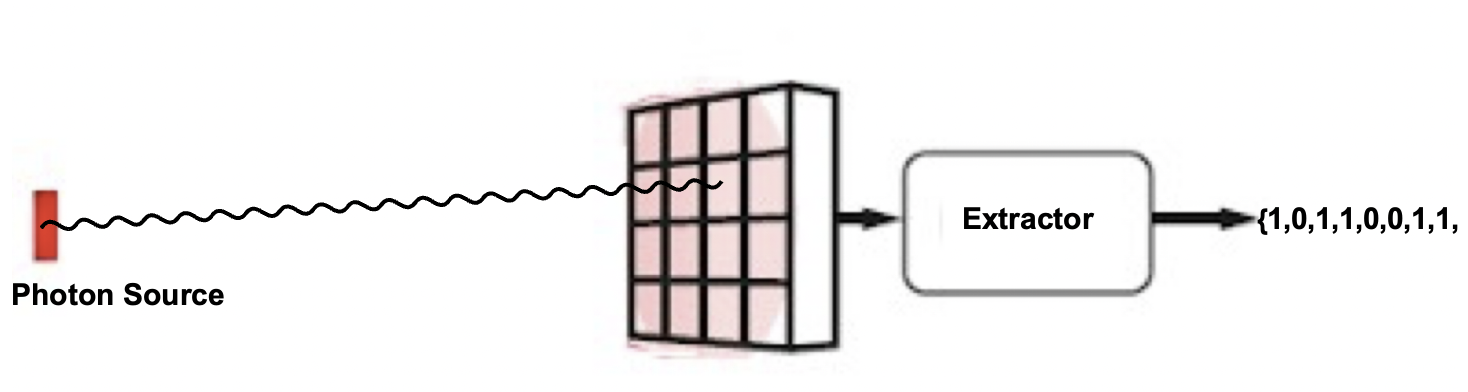}
  \caption{Quantum random number generator}
  \label{fig:random}
  \end{figure}
What one can predict is the probability of the particle arriving somewhere but arriving at some fixed spot is
something which, from the principles of quantum mechanics, cannot be reproduced. This means that there is some
fundamental randomness which is inherent to quantum physics and has the potential of providing true random numbers. It thus has become the preferred option for the scientific applications requiring randomness (Figure \ref{fig:random}).

On the mathematical side of things, there is a related question  that in fact goes back to Newton, which is simply \emph{what is a real variable?}. If you ask a mathematician for his view
 on this question the most likely answer you will get is that \emph{a real variable is given by a set $X$ and a map from this set $X$ to
the real numbers}. However, if you think more deeply about it, you will find out that this answer is very unsatisfactory. Indeed, it means that one can not have coexistence of continuous variables, namely variables which can take a continuous range of possible values, and discrete variables, namely
variables which can only affect a countable set of possible values, say, with finite multiplicity for each of them.
The reason for this is that if you have a discrete variable then the original set $X$ that you are dealing with will have to
be countable, which then implies that it does not allow for continuous variables.

The amazing answer that mathematics provides, but which would
not have been detected if it were not for the formalism of quantum mechanics by von Neumann, is that a real variable is just a self-adjoint operator in a Hilbert space. As a matter of fact,  there is only one (separable) Hilbert space which is the Hilbert space with countable basis and one sees  
that this Hilbert space has variables with discrete spectrum. Indeed, take a description of Hilbert space by giving a countable orthogonal basis
and take an operator which is diagonal in that basis. This operator has countable spectrum but it coexists with operators which have continuous spectrum. In fact one could have described that same 
Hilbert space as being the space of square integrable functions on an interval and of course one will have continuous variables there, given by 
multiplication operators. The beauty of this formalism is that {\em continuous and discrete variables do coexist}. All the properties of real variables are there because if one has a self-adjoint operator it has a spectrum which is composed of the possible values of the variable and it has spectral multiplicity which is the number of times a value can be affected.

\begin{table}
\begin{tabular}{r | l}
& \\
 Real variable $f:X\to \R$  & Self-adjoint operator $H$ in Hilbert space  \\
Range $f(X)\subset \R$ of the variable & Spectrum of  the operator $H$\\ 
Composition $\phi\circ f$, $\phi$ measurable & Measurable functions $\phi(H)$ of self-adjoint operators  \\
Bounded complex variable $Z$ & Bounded operator $A$ in Hilbert space \\
Infinitesimal variable $dx$ & Compact operator $T$\\
Infinitesimal of order  $\alpha>0$  & Characteristic values $\mu_n(T)=O(n^{-\alpha})$ for $n\to \infty$ \\
 Algebraic operations on functions & Algebra of operators in Hilbert space \\ 
Integral of function $\int f(x)dx$ & $\displaystyle{\int\!\!\!\!\!\! -} T =$ coefficient of $\log(\Lambda)$ in $\Tr_\Lambda(T)$\\
Line element $ds^2=g_{\mu\nu} dx^\mu dx^\nu$ & $ds=\bullet\!\!\!\!-\!\!\!-\!\!\!-\!\!\!-\!\!\!-\!\!\bullet$ : Fermion propagator  $D^{-1}$\\
$d(a,b)=\,{\rm Inf}\,\int_\gamma\,\sqrt{g_{\mu\,\nu}\,dx^\mu\,dx^\nu}$ & $d(\mu,\nu)=\,{\rm Sup}\,\vert \mu(A)-\nu(A)\vert, \mid \ \Vert [D,A]\Vert\leq 1.$ \\ 
Riemannian geometry $(X,ds^2)$& Spectral geometry $(\cA,\cH,D)$\\
Curvature invariants & Asymptotic expansion of spectral action \\
Gauge theory & Inner fluctuations of the metric\\
&
\end{tabular}
\caption{The `spectral' dictionary}
\label{spectral-dictionary}
\end{table}

All this fits very well with reality in the sense that the quantum variables are operators in 
Hilbert space; the new key fact is that the discrete variables do not commute with the continuous variables. If they would commute, they would both be functions defined on the same space $X$, which is not possible.

Summarizing, continuous and discrete variables  coexist as operators in Hilbert space, but as such they necessarily {\em do not commute}. It is precisely this lack of commutativity which is the new ingredient at the core of the quantized calculus   and which renders the framework of non-commutative
geometry effective.

\section{Spectral paradigm of geometry, the line element}
           \label{sectlocalact}
           
What has happened in the process which led to understand the emergence of geometry from the quantum  is quite  instructive. When Heisenberg found his commutation relations involving $P$
and $Q$ (momentum and position) there was already quite a hint, a bit of truth, in it, in the sense 
 that when you take the spectrum of either $P$ or $Q$ you find a real line. The other operator is  a  differentiation operator and it gives
  a geometrical structure for that line. However, the way things evolved historically from  these commutation relations, 
 is that they were interpreted not as a first hint of "geometry from Hilbert space operators" but rather in terms of  Lie group representations. Of course, group representations, when applied to {\em e.g.} the Poincar\' e group give a beautiful conceptual notion of particle and the theory unveiled big pieces of beautiful landscape. However, finite-dimensional Lie groups can hardly lead us to the arbitrary geometries that we observe in gravity and to the variables that we have in gravity. Simple Lie groups are like isolated diamonds but which simply do not allow for this enormous variability needed in the geometrical description of gravity.

The development of noncommutative geometry has shown that it is possible to
have geometry emerging from purely Hilbert space considerations but in order to do that one applies representation theory to far more elaborate forms of  Heisenberg's commutation relations involving $P$ and $Q$.   In order to obtain these relations one needed to take a step back 
and understand how to give more flexibility to both $P$ and $Q$. The additional flexibility for $Q$ was quite hard to come by and is discussed in Section \ref{sectalgebra} below. The additional flexibility for $P$ was not so hard to find. In fact, it was
found already by Paul Dirac in \cite{Dir27} when he realised how to assemble several momenta together to form a single expression, a single operator,  that actually contains
in itself all the components of the momenta. This is the {\em Dirac operator}, expressed in terms of gamma matrices. Thus the understanding of how to give more flexibility to $P$ stems from Dirac's work and, even more, this understanding is thoroughly grounded in physics and in the understanding of geometry, in particular in the measurement of lengths, as we will now explain. 

Many formalisms of geometry start with the Riemannian paradigm as a prerequisite, that is, the idea that geometry is given by the 
measurement of lengths and this measurement of lengths is actually governed locally by simply prescribing the square of the
line element $ds^2=g_{\mu\nu}dx^\mu dx^\nu$. It turns out that this idea was even questioned by Riemann himself!
He wrote in his inaugural lecture on the foundation of geometry  that it's questionable whether the texture of space (or spacetime) will obey this ``Riemannian paradigm" at any scale, the reason being that the notions of light ray and of solid body on which his intuition was grounded would cease to make sense for very tiny scales.

At  the end of the
18th century the desire to unify the measurement of distances led to a concrete realization of a unit of length which was called ``m\`{e}tre \'{e}talon" and was conserved near Paris in the form of a platinum bar. Later, 
the relevance of this choice was put into question in the early 1920's because people found out that the m\`{e}tre \'{e}talon 
was actually changing length which of course was very problematic. They found this out because they 
measured the m\`{e}tre \'{e}talon by comparing it with the wavelength of a fixed atomic transition of Krypton. The outcome of this observation,  many
years later,  was that physicists shifted the definition of the unit of length from a platinum metal bar to the  wavelength of 
a certain transition of Krypton and later, eventually, to an hyperfine  transition of Caesium. 

If you think more deeply about this, you find that the reason why their classical unit of length needed to be localized was because first it should be quite small, since it is supposed to represent $ds$, and because it commuted with 
the coordinates it had to be localized somewhere (as it happened this was near Paris). However, the new unit of length which is given by a wavelength of the hyperfine transition of Caesium\footnote{It is called a unit of time using the speed of light as  the conversion factor}, is in fact of spectral nature and is no longer commuting with the coordinates. 
As a matter of fact, in line with the above discussion, it involves the Dirac operator, or, rather its inverse, the Dirac propagator. Obviously, because it does not commute with the coordinates, it does not need to be localised. Moreover, if one wants to unify 
the metric system in our galaxy it is clear that such a spectral definition should be used as a unit of length. Indeed, it is much more practical to tell people from nearby stars that our unit of length is  a certain transition in the helium or hydrogen spectrum rather than to tell them that they need  to come to Paris and compare their unit with the metal bar which is located there.

Besides the physical standpoint that this is a much better definition of the unit of length, mathematically speaking it implies that one actually replaces the $ds^2$ of Riemannian geometry by a very subtle square root of it, where the square root is taken through the
Clifford algebra and where the infinitesimal line element which was formulated in terms of infinitesimal
variables by Riemann is replaced by an infinitesimal which is a suitable operator in Hilbert space (see Table \ref{spectral-dictionary}). 
Note that this notion of infinitesimal was in fact predicted by Newton in the sense that
 he said explicitly that an infinitesimal should not be a number but instead it 
should be a variable! In fact, Newton gave the definition of an infinitesimal variable and, when you translate it in terms of the understanding of variables using Hilbert space
operators, it gives precisely what we call a compact operator. These compact operators have exactly all the properties you would dream of  for infinitesimals, they
form an ideal among operators etc. etc...

Now, when you think about the line element for spacetime, you find out that the 
mathematical formulation consisting of the inverse of the Dirac
operator, in the language of physics, is encoded by what is usually referred to as the the fermion propagator. This fermion
propagator enters Feynman graphs as the internal legs involving fermions. When you look 
at quantum field theory textbooks \`{a} la Feynman, you find that this propagator is a small tiny
line joining two very nearby events:
\begin{center}
  \includegraphics[scale=.05]{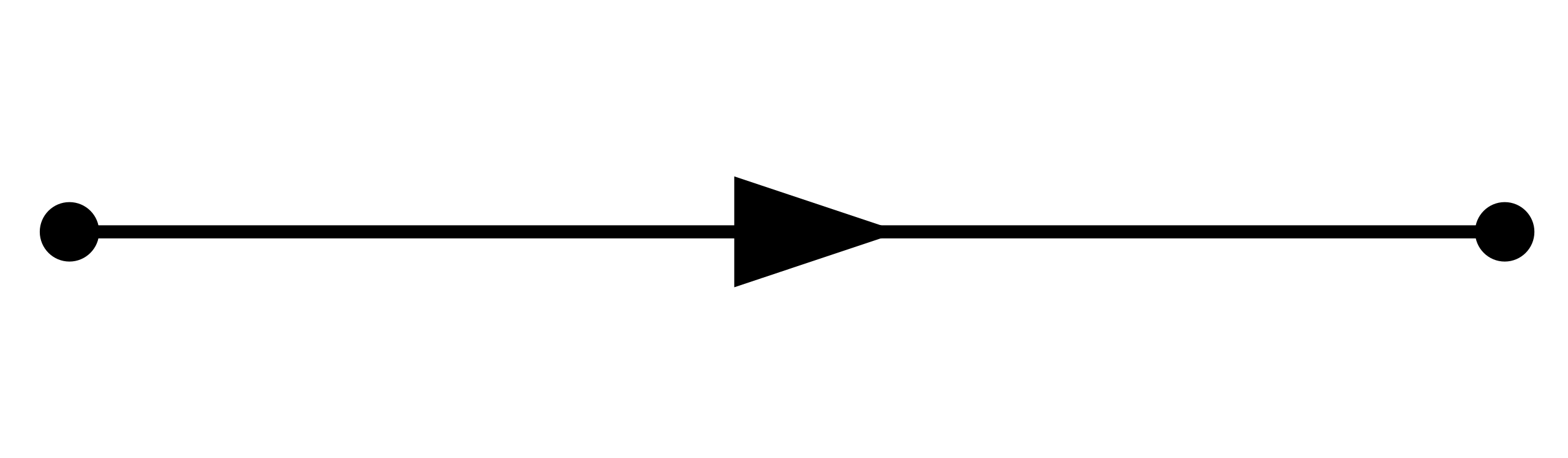}
  \end{center}
This qualitative appearance of an infinitesimal line element becomes even more striking when one realizes that in quantum field theory, 
 this line element actually acquires quantum corrections by being dressed. This means that
if you take from the beginning the correct version of the line element for geometry, you soon understand that
there are quantum corrections to the measurement of lengths, to the geometry, which are given by the dressing of the fermion
propagator. Moreover, gauge fields appear as additional terms in the Dirac operator and this embodies exactly the intuition of Riemann that in case his paradigm would fail at small distances the geometry should be based on the forces that hold things together!

\section{Spectral paradigm of geometry, the algebra}\label{sectalgebra}

The 
 new spectral paradigm of geometry is encoded by {\em Spectral triples} $(\cA,\cH,D)$ where the unbounded operator $D$ encodes the analogue of the Dirac operator as described in Section \ref{sectlocalact} and where $\cA$ is an algebra of operators in the Hilbert space $\cH$. The algebra $\cA$ is not assumed to be commutative and  its inner automorphisms (as explained in Section \ref{secttime}) will correspond to the internal symmetries in physics.  Because of its flexibility, this new paradigm provides the needed tool to 
 refine our understanding of the structure of physical space in the small and to ``{\it seek the foundation of its metric relations outside it, in binding
forces which act upon it}". The main idea, described in details in \cite{CK}, is that the line element now embodies not only the force of gravity but all the known forces,  electroweak and strong, appear from the spectral action and the inner fluctuations of the metric. This provides   a completely new perspective on the geometric interpretation of the detailed  structure of the Standard model and of the Brout-Englert-Higgs mechanism. One gets the following simple mental picture for the appearance of the scalar field: imagine that the space under consideration is two sided like a sheet $S$ of paper in two dimensions. Then when differentiating a function on such a space one may restrict the function to either side $S_\pm$ of the sheet and thus obtain two spin one fields. But one may also take the finite difference $f(s_+)-f(s_-)$ of the function at the related points of the two sides. The corresponding field is clearly insensitive to local rotations and is a scalar spin zero field. This, in a nutshell, is how the Brout-Englert-Higgs  field appears geometrically once one accepts that there is a ``fine structure" which is revealed by the detailed structure of the standard model of matter and forces. Eventually, this allows one to uncover  the  
 geometric meaning of the Lagrangian of gravity coupled to the standard model. This extremely complicated Lagrangian is obtained from the spectral action developed in \cite{cc2} which is the only natural additive spectral invariant of a noncommutative geometry.

 In order to comply with Riemann's requirement that the inverse line element $D$ embodies the forces of nature, it is evidently important that we do not separate artificially the gravitational part from the gauge part, and that $D$ encapsulates both forces in a unified manner. In the traditional geometrization of physics, the gravitational part specifies the metric while the gauge part corresponds to a connection on a principal bundle. In the noncommutative geometric framework, $D$ describes both forces in a unified manner and the gauge bosons appear as inner fluctuations of the metric but form an inseparable part of the latter. The key point is that {\em non-abelian} gauge fields arise inevitably as a result of the {\em noncommutativity} of our geometric framework.

 The noncommutative geometry dictated by physics is then given by the product of the ordinary $4$-dimensional continuum by a finite noncommutative geometry $(\cA_F,\cH_F,D_F)$ which appears naturally from the classification of finite geometries of $KO$-dimension  equal to $6$ modulo $8$ (\cf \cite{cc5,mc2}). The finite dimensional algebra $\cA_F$ which appeared is of the form 
$$
\cA_F=C_{+}\oplus C_{-}; \qquad  C_{+}=M_{2}(\mathbb{H}),\  \  \  C_{-}=M_{4}(\mathbb{C}).
$$

The agreement of the mathematical formalism of spectral geometry and all its subtleties such as the periodicity of period $8$  of the $KO$-theory, is promising but might still be accidental. Instead, what is most convincing is the pertinence of this approach when recovering \cite{mc2} the see-saw mechanism (which was dictated by the pure math calculation of the model), while being unaware at the time of its key physics role to provide very small non-zero masses to the neutrinos, and of how it is ``put by hand" in the standard model. The low Higgs mass then came in 2012 as a possible flaw of the model, but in \cite{acresil} we showed the compatibility of the model with the measured value of the Higgs mass, due to the role in the renormalization of the scalar field which was already present in \cite{ac2010} but had been ignored thinking that it would not affect the running of the self-coupling of the Higgs.

In all the previous developments we had followed the ``bottom-up'' approach, \ie we uncovered the details of the finite noncommutative geometry $(\cA_F,\cH_F,D_F)$ from the experimental information contained in the standard model coupled to gravitation. In 2014, 
in collaboration with S. Mukhanov \cite{acmu1,acmu2} we were investigating  the purely geometric problem of encoding $4$-manifolds in the most economical manner in the spectral formalism. The problem had no a priori link with the standard model of particle and forces and the idea was to treat the coordinates in the same way as the momenta are assembled together in a single operator using the gamma matrices. The great surprise was that this investigation  gave the conceptual explanation of the finite noncommutative geometry from Clifford algebras! This is described in details in \cite{CK} to which we refer. What we  obtained is a higher form of the Heisenberg commutation relations between $P$ and $Q$, whose irreducible Hilbert space representations correspond to $4$-dimensional spin geometries. The role of $P$ is played by the 
Dirac operator as explained before, and the role of $Q$ by the Feynman slash of coordinates using Clifford algebras. The proof that all spin geometries are obtained relies on deep results of immersion theory and ramified coverings of the sphere.  The volume of the $4$-dimensional geometry is automatically quantized by the index theorem and the  spectral model, taking into account the inner automorphisms due to the noncommutative nature of the Clifford algebras, gives Einstein gravity coupled with the slight extension of the Standard Model, as a Pati--Salam model. This model was shown in \cite{acpati1,acpati2} to yield unification of coupling constants. We refer to the survey \cite{chamssuijl} for a concise account of the whole story of the evolution of this theory from the early days to now. 

\begin{table}[H]
\begin{center}
\begin{tabular}{r | l}
  {\bf Standard Model} & {\bf Spectral Model}\\
  \hline
 Higgs Boson &  Inner metric$^{(0,1)}$ \\
   Gauge bosons  &  Inner metric$^{(1,0)}$ \\
     Fermion masses $u,\nu$&     Dirac$^{(0,1)}$ in $\uparrow$\\
    CKM matrix,  Masses down &   Dirac$^{(0,1)}$ in $\dodo$\\
  Lepton mixing, Masses leptons $e$ &   Dirac$^{(0,1)}$ in $\elel$\\
         Majorana  mass matrix & Dirac$^{(0,1)}$ on $E_R\oplus J_F E_R$\\
  Gauge couplings  & Fixed at unification\\
  Higgs scattering parameter   &  Fixed at unification  \\
  Tadpole constant  &  $- \mu_0^2\, |\higgs|^2$\\
  \end{tabular}
\end{center}
\caption{The dictionary between the physics terminology and the fine structure of the geometry}
\end{table}

\section{Dimension $4$}\label{sectsobolev}
The above approach to the geometry of space-time was not directly motivated by quantum gravity but it addresses a more basic preliminary question which is to understand the reason why gravity together with the standard model appear as the fundamental forces of nature. We have seen in Section \ref{sectalgebra} that as long as one allows the encoding of the   algebra of coordinates using a minimal amount of noncommutativity, gravity coupled to the standard model appears naturally with the non-abelian gauge theories as witnesses of the noncommutative encoding. We now discuss briefly how dimension $4$ is singled out in this approach.  As explained above the   extension of Riemannian geometry beyond its classical domain, in the framework of Hilbert space operators,  provides the needed flexibility in order to answer the query of Riemann in his inaugural lecture. From the mathematical standpoint it is the very notion of manifold which is at centerstage, and when one wants to capture the key properties fulfilled by manifolds at a global (non-local) level, one finds that the main one is Poincar\'e duality. But it is not Poincar\'e duality in ordinary homology but rather in a more refined theory called $KO$-homology. A remarkable interaction with the quantum formalism then appears  from the confluence of the abstract understanding of the notion of manifold from its fundamental class in $KO$-homology with  the realization of cycles in $KO$-homology from representations in Hilbert space.  The final touch on the understanding of the geometric reason behind gravity coupled to the standard model, comes from the simultaneous quantization of the fundamental class in $KO$-homology and its dual in $KO$-theory which gives rise to the  higher Heisenberg relation. We now explain in more details why dimension $4$ plays a special role in this context \cite{acmu1,acmu2}. 
 In order to encode a manifold $M$ of dimension $d$ using the above duality between $KO$-homology and  $KO$-theory one needs to construct a pair of maps $\phi,\psi$ from $M$  to the sphere (of the same dimension $d$) in such a way that the sum of the pullbacks
of the volume form of the (round) sphere vanishes nowhere on $M$. This problem is easy to solve in dimension $2$ and $3$ because one first writes $M$ as a ramified cover $\phi:M\to S^d$  of the sphere and one pre-composes   $\phi$ with a diffeomorphism $f$ of $M$ such that $\Sigma\cap f(\Sigma)=\emptyset$ where $\Sigma$ is the subset where $\phi$ is ramified. This subset is of codimension $2$ in $M$ and there is no difficulty to find $f$ because $(d-2)+(d-2)<d$ for $d<4$. It is worth mentioning that the $2$ for the codimension of $\Sigma$ is easy to understand from complex analysis: For an arbitrary smooth map $\phi:M\to S^d$ the Jacobian will vanish on a codimension $1$ subset, but in one dimensional  complex analysis the Jacobian is a sum of squares and its vanishing means the vanishing of the derivative which gives two conditions rather than one.   Thus {\bf dimension $d=4$ is the critical dimension} for the above existence problem of the pair $\phi,\psi$. Such a pair does not always exist\footnote{It does not exist for $M=P_2(\C)$.} but as shown in \cite{acmu1,acmu2}, it always exist for spin manifolds which is the relevant case for the functional integral performed in Euclidean signature. 

The higher form of the Heisenberg commutation relation mentioned above involves in dimension $d$ the power $d$ of the commutator $[D,Z]$ of the Dirac operator with the operator $Z$ which is constructed (using the real structure $J$, see \cite{CK}) from the coordinates. We shall now explain briefly how this fits perfectly with the framework of D. Sullivan on Sobolev manifolds, \ie of manifolds of dimension $d$ where the pseudo-group underlying the atlas  preserves continuous functions with one derivative in $L^d$.   He discovered  the intriguing special role of dimension $4$ in this respect. He showed in \cite{S3} that 
 topological manifolds in dimensions $>5$ admit bi-Lipschitz coordinates and these are unique up to small perturbations, moreover existence and uniqueness also holds for  Sobolev structures: one derivative in $L^d$.
A stronger result was known classically for dimensions 1,2 and 3. There the topology controls the smooth structure up to small deformation. 
In dimension 4, he proved with S. Donaldson in \cite{DS} that for manifolds with coordinate atlases related by the  pseudo-group  preserving continuous functions with one derivative in $L^4$ it is possible to  develop the $SU(2)$ gauge theory 
and the  famous Donaldson invariants. Thus in dimension $4$ the Sobolev manifolds  behave like the smooth ones as opposed to Freedman's abundant topological manifolds.\footnote{for which  any modulus of continuity  whatsoever is not known.} The obvious question then is to which extent  the higher Heisenberg equation of \cite{acmu1,acmu2} singles out the Sobolev manifolds as the relevant ones for the functional integral involving the spectral action. 


\section{The second quantized level}\label{secondq}
This final section is more speculative and unlike the previous ones it addresses a fundamental question which is essentially open. This question can be formulated as follows 

\begin{quote}
{\em What is the relevance of the many particle formalism of Quantum Field Theory to the geometry of space-time?}
\end{quote}
In fact we have already seen a hint of an answer in Section \ref{sectlocalact} when we pointed out that since the line element encoded by the Dirac propagator gets dressed  (as a formal power series in powers of $\hbar$) from the quantum field corrections, this suggested that the geometry itself, being encapsulated by the Dirac propagator gets dressed.  
But the above formalism of geometry remained at the level of first quantization and the issue of second quantization  cannot be ignored since the quantum corrections to the line element as explained in Section \ref{sectlocalact} above are only the tip of the iceberg. Indeed, the dressing occurs for all the $n$-point functions for fermions, and not only the two point function.\newline
Another significant hint for the role of the second quantization is the result of \cite{CCS} on the second quantization  of fermions and the spectral action as an entropy, The point there is that the action principle which recovers gravity coupled with the Standard Model is given by the most general additive functional of spectral geometries and these depend on an a priori arbitrary even function $f$ of a real variable which defines the action as the trace of $f(D)$. What we showed in \cite{CCS}, as a first step towards building from quantum field theory a {\em second quantized} version of spectral geometry, is that if one applies the free functorial quantization to Fermions as Clifford algebras and uses the Dirac operator to define a time evolution on this Clifford algebra one discovers that the von Neumann entropy of the Boltzmann-Gibbs equilibrium states (see Section \ref{secttime}) is the spectral action for a function $f$ deeply related to the Riemann zeta function. This led us to the first lines of a dictionary from first to second-quantized as given in Table \ref{2nd-quant-dictionary}. Note also that an analogous result to \cite{CCS} has been obtained in \cite{DK} for the bosonic case.

\begin{table}[H]
  \begin{center}
    \begin{tabular}{r | l}
     {\bf First-quantized} & {\bf Second-quantized}\\
      \hline
  Algebra & Action of inner fluctuations on time evolution\\
  Hilbert space & Clifford algebra\\
  Dirac operator& Time evolution for KMS$_{\beta}$-state on Clifford algebra\\
  Spectral action & Entropy of KMS$_\beta$-state
\end{tabular}
  \end{center}
  \caption{The dictionary for fermionic second-quantization of spectral triples}
          \label{2nd-quant-dictionary}
  \end{table}
 From the purely mathematical standpoint this need   to pass to a second quantized higher level of geometry is in a way similar to what has happened in the development of $K$-theory. The topological $K$-theory as developed by Atiyah-Hirzebruch based on Bott periodicity leads  to the key  duality between $KO$-homology and $KO$-theory is the origin of the higher Heisenberg relation. As already mentioned in \cite{coinaugural}, algebraic $K$-theory, which is a vast refinement of topological $K$-theory,  is begging for the development of a dual theory and one should expect profound relations between this dual theory and the  theory of interacting quanta of geometry. As a concrete point of departure, note that the deepest results on the topology of diffeomorphism groups of manifolds are given by the Waldhausen algebraic $K$-theory of spaces and we refer to \cite{DGM} for a unifying picture of algebraic $K$-theory. 	\newline
 Finally, the conceptual understanding of renormalization has shown (see \cite{CMbook})
that there is a highly noncommutative symmetry group, the {\em Cosmic Galois group}, which embodies the ambiguities in the renormalization process inherent to the physics computations of interacting Quantum Field Theory. How this symmetry group combines with the above impact of physics on geometry, and how renormalization fits with the spectral action principle are deep open questions of which only the first few steps have recently been taken \cite{NS21,NS21b}. These questions are witnesses of the amazing role of noncommutativity in physics.

\end{document}